\documentclass[11pt,prd, aps, showpacs,preprint,superscriptaddress]{revtex4}
\usepackage[pdftex]{graphicx,color}
\usepackage[english]{babel}
\usepackage{amsmath}
\usepackage{jan}
\newcommand\edinb{SUPA, School of Physics and Astronomy, The University of
  Edinburgh, Edinburgh EH9 3JZ, UK}

\begin{document}
\title{The S Parameter in QCD from Domain Wall Fermions}
\author{Peter A. Boyle}
\author{Luigi Del Debbio}
\author{Jan Wennekers}\email{jwenneke@ph.ed.ac.uk}
\author{James M. Zanotti}
\affiliation{\edinb}
\preprint{Edinburgh 2009/11} \collaboration{RBC and UKQCD collaborations}
\date{\today} 
\pacs{11.15.Ha, 
      11.30.Rd, 
      12.38.Gc  
      12.39.Fe  
}
\begin{abstract}
We have computed the $SU(2)$ Low Energy Constant $l_5$ and the mass
splitting between charged and neutral pions from a lattice QCD
simulation of $n_f=2+1$ flavors of Domain Wall Fermions at a scale of
$a^{-1}=2.33\,\text{GeV}$. Relating $l_5$ to the $S$ parameter in QCD
we obtain a value of $S(m_H=120\,\text{GeV})=0.42(7)$, in agreement
with previous determinations. Our result can be compared with the
value of $S$ from electroweak precision data which constrains strongly
interacting models of new physics like Technicolor. This work in QCD
serves as a test for the methods to compute the $S$ parameter with
Domain Wall Fermions in theories beyond the Standard Model. We also
infer a value for the pion mass splitting in agreement with experiment.
\end{abstract}
\maketitle

\section{Introduction}
Models of dynamical electroweak symmetry breaking (DEWSB) like
Technicolor are possible extensions of the Standard Model, and many
proposals have been put forward for a strongly--interacting sector
beyond the Standard Model since the original proposals in
Refs.~\cite{Weinberg:1979bn,Susskind:1978ms}. Viable models of DEWSB
must satisfy the constraints that follow from electroweak precision
data~\cite{Peskin:1990zt,Altarelli:1990zd}. These constraints put
severe limitations on Technicolor candidates, and QCD--like theories
naively rescaled to the electroweak scale are already ruled out, see
e.g. Refs.~\cite{Hill:2002ap,Sannino:2008ha} for recent reviews.  The
constraints are nicely encoded in bounds for the value of the $S$
parameter introduced in Refs.~\cite{Peskin:1990zt,Peskin:1991sw}.

Theories near an infrared fixed point (IRFP) have been advocated as
promising candidates for DEWSB.  Preliminary attempts at estimating
analytically the $S$ parameter for these theories suggest that it is
much reduced compared to QCD--like
theories~\cite{Appelquist:1998xf,Harada:2005ru}. Unfortunately these
computations are not based on first principles, and have to rely on
assumptions that are difficult to control. Early studies focused on
finding evidence for an IRFP in theories with a large number of
flavors in the fundamental representation of the gauge group
following the seminal example in Ref.~\cite{Banks:1981nn}. More
recently, novel models have been proposed based on smaller numbers of
fermions in higher--dimensional
representations~\cite{Dietrich:2006cm}. Several phenomenological
scenarios have been proposed which build upon these
ideas~\cite{Luty:2004ye,Foadi:2007ue}.

Lattice simulations are now in a position to address these questions
from first principles, so that the difficulties in dealing with the
non-perturbative dynamics can be dealt with in a systematic way. The
existence of IRFPs has been investigated in theories with fundamental
fermions~\cite{Appelquist:2007hu,Deuzeman:2008sc,
Deuzeman:2009mh,Fodor:2009nh,Hasenfratz:2009ea,Fodor:2009wk,Appelquist:2009ty}
and higher
representations~\cite{Catterall:2007yx,DelDebbio:2008wb,Shamir:2008pb,
DelDebbio:2008zf,Catterall:2008qk,DeGrand:2008kx,Hietanen:2008mr,
Hietanen:2009az,DeGrand:2009mt,DeGrand:2009et,DelDebbio:2009fd}. These
preliminary studies have mapped out the space of bare lattice
parameters and are starting to study the spectrum of the candidate
theories, and the flow of renormalized couplings. First results hint
towards an interesting landscape of theories that could exhibit scale
invariance at large distances.

Computing the $S$ parameter in these theories from first principles is
an important ingredient in trying to build successful phenomenological
models. The $S$ parameter is obtained on the lattice from the form
factors that appear in the momentum--space VV-AA correlator, as
defined below in Section~\ref{sec:vac}.  Chiral symmetry plays an
important role in guaranteeing the cancellation of power--divergent
singularities when computing the above correlator. Hence lattice
formulations that preserve chiral symmetry at finite lattice spacing
are particularly well--suited for these studies. QCD is the ideal
testing ground to develop and test the necessary lattice
technology. In this work, we compute the $S$ parameter in QCD with
$n_f=2+1$ flavors of Domain Wall fermions (DWF). Our study closely
follows the procedure described in Ref.~\cite{Shintani:2008qe} where
the $S$ parameter (or equivalently the $SU(3)$ Low Energy Constant $L_{10}$)
was first computed from vacuum polarisation functions (VPFs) using
overlap fermions. We adopt the method, apply it to Domain Wall
Fermions, and widen its scope by using conserved currents and larger
physical volumes.

Section~\ref{sec:vac} contains a short account of computational
methods for the VPF on the lattice. Section~\ref{sec:contact}
addresses the topics of power divergences and residual chiral
symmetry breaking. In Section~\ref{sec:res} we present the numerical
results obtained on the gauge configurations produced by the RBC and
UKQCD collaborations using Domain Wall
Fermions. Section~\ref{sec:pion} contains our results for the pion
mass splitting. A discussion of the numerical results and a short
conclusion can be found in Section~\ref{sec:conc}.

\section{Vacuum Polarisation Functions}
\label{sec:vac}
Domain Wall Fermions are a five dimensional formulation of lattice QCD
with an approximate chiral symmetry \cite{Kaplan:1992bt,Shamir:1993zy,
Furman:1994ky}. The residual explicit breaking of chiral symmetry
appears in the Ward identities of the theory as terms proportional to
the so--called residual mass $m_\mathrm{res}$. The conserved vector
and axial currents form a multiplet under this approximate lattice
chiral symmetry. 

The basic observables in this work are vacuum polarisation functions of
the vector and the axial vector current. They are defined as
current-current two-point functions in momentum space,
\begin{align}
  \Pi_{\mu\nu}^V(q)&\equiv\sum_{x}\e^{\ii q\cdot x}
  \vev{0|\mathcal{V}_\mu(x)V_\nu(0)|0},\label{eq:piv}\\
  \Pi_{\mu\nu}^A(q)&\equiv\sum_{x}\e^{\ii q\cdot x}
  \vev{0|\mathcal{A}_\mu(x)A_\nu(0)|0},\label{eq:pia}
\end{align}
where $\mathcal{V}_\mu$ and $\mathcal{A}_\mu$ are the conserved vector
and axial currents and $V_\mu$ and $A_\mu$ are the corresponding local
currents.  We consider local-conserved correlators from a new set of
point source propagators with up to two units of spatial momentum. The
definition of the conserved currents can be found in
Ref.~\cite{Furman:1994ky}. Since the Fourier transform in
Eqs.~(\ref{eq:piv},\ref{eq:pia}) includes $x=0$, power--divergent
contributions can arise. Due to lattice chiral symmetry these
divergences cancel in the difference of the vector and axial vector
correlators, if conserved currents are used. Note that similar
observables are used to compute hadronic contributions to the
anomalous magnetic moment of the muon in Refs.~\cite{Blum:2002ii,
Gockeler:2003cw, Aubin:2006xv}, where one can find a more detailed
discussion of the renormalization of the correlators. A similar
cancellation was pointed out in Ref.~\cite{Shintani:2008qe}, where
overlap fermions have been used.

Following Ref.~\cite{Shintani:2008qe} we decompose the difference
$\Pi_{\mu\nu}^{V-A}\equiv\Pi_{\mu\nu}^{V}-\Pi_{\mu\nu}^A$ into a
longitudinal and a transverse part,
\begin{align}
  \label{eq:pi1}
  \Pi_{\mu\nu}^{V-A}=\left(q^2\delta_{\mu\nu}-q_\mu q_\nu\right)
  \Pi^{(1)}(q^2) - q_\mu q_\nu\Pi^{(0)}(q^2). 
\end{align}
For each momentum we average all components of $\Pi_{\mu\nu}$ which
contribute to only one of the ``polarisations''. For example for
momenta with one non-zero direction $q_\kappa$ we have
$q^2\Pi^{(0)}=\Pi_{\kappa\kappa}^{V-A}$ and 
$q^2\Pi^{(1)}=\frac{1}{3}\sum_{\mu\neq\kappa}\Pi_{\mu\mu}^{V-A}$.

We use $SU(2)$ Chiral Perturbation Theory (ChPT) to fit to our data to
 extract the Low Energy Constant $l_5^r$. Since we are using $2+1$
 flavor lattices the low-energy constants will implicitly depend on
 the strange quark mass. The more common notation in the literature is
 to use the $SU(3)$ LEC $L_{10}^r$ which is related to $l_5^r$ by
\begin{align}
  \begin{split}
  L_{10}^r&=l_5^r-\frac{1}{384\pi^2}\left(\log\frac{m_K^2}{\mu^2}+1\right),\\
  &=l_5^r-3\cdot10^{-5}\;\text{for}\;\mu=m_\rho.
  \end{split}
\end{align}
The ChPT result for $\Pi^{(1)}$ can be found in Ref.~\cite{Gasser:1984gg}:
\begin{align}
  q^2\Pi^{(1)}\left(m_\pi,q^2\right)&=
  -f_\pi^2-\left[\frac{1}{48\pi^2}\left(\bar{l}_5-\frac{
      1}{3}\right)
    -\frac{1}{3}\sigma^2\bar{J}(\sigma)\right]q^2
    +O(q^4)\label{eq:chpt},\\
    \text{with}\quad\bar{J}(\sigma)&=\frac{1}{16\pi^2}\left
    (\sigma\log\frac{\sigma-1}{\sigma+1}+2\right)\quad\text{and}\quad
    \sigma=\sqrt{1-\frac{4m_\pi^2}{q^2}}.
\end{align}
With $m_\pi$, $f_\pi$ known from fits to pseudoscalar and $A_0$
correlators (Table \ref{tab_meson}) there is only one free parameter
in (\ref{eq:chpt}) for each choice of the chiral scale $\mu$. The scale
invariant LEC $\bar{l}_5$ is defined by
\begin{align}
  \bar{l}_5=-192\pi^2\,l_5^r(\mu)-\log\frac{m_\pi^2}{\mu^2}, 
\end{align}
and the corresponding convention for the $S$ parameter
\cite{Peskin:1990zt} is
\begin{align}
  S=\frac{1}{12\pi}\left[-192\pi^2l_5^r(\mu)+\log(\mu^2/m_H^2)-\frac{1}{6}\right].
  \label{eq:sl5}
\end{align}

\section{Contact Terms in Ward Identities}
\label{sec:contact}
\begin{figure}[t]
  \includegraphics[width=.45\textwidth]{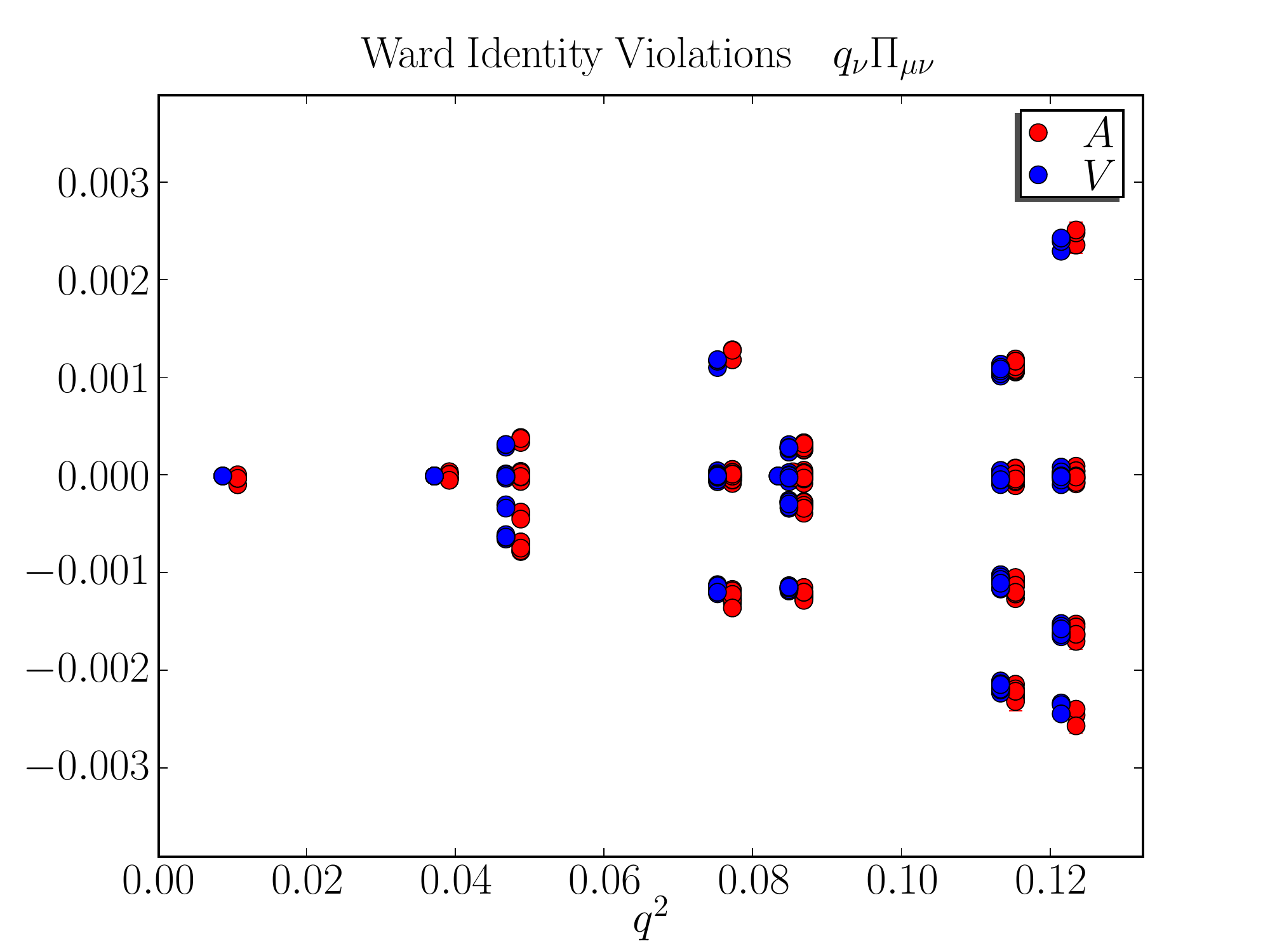}
  \includegraphics[width=.45\textwidth]{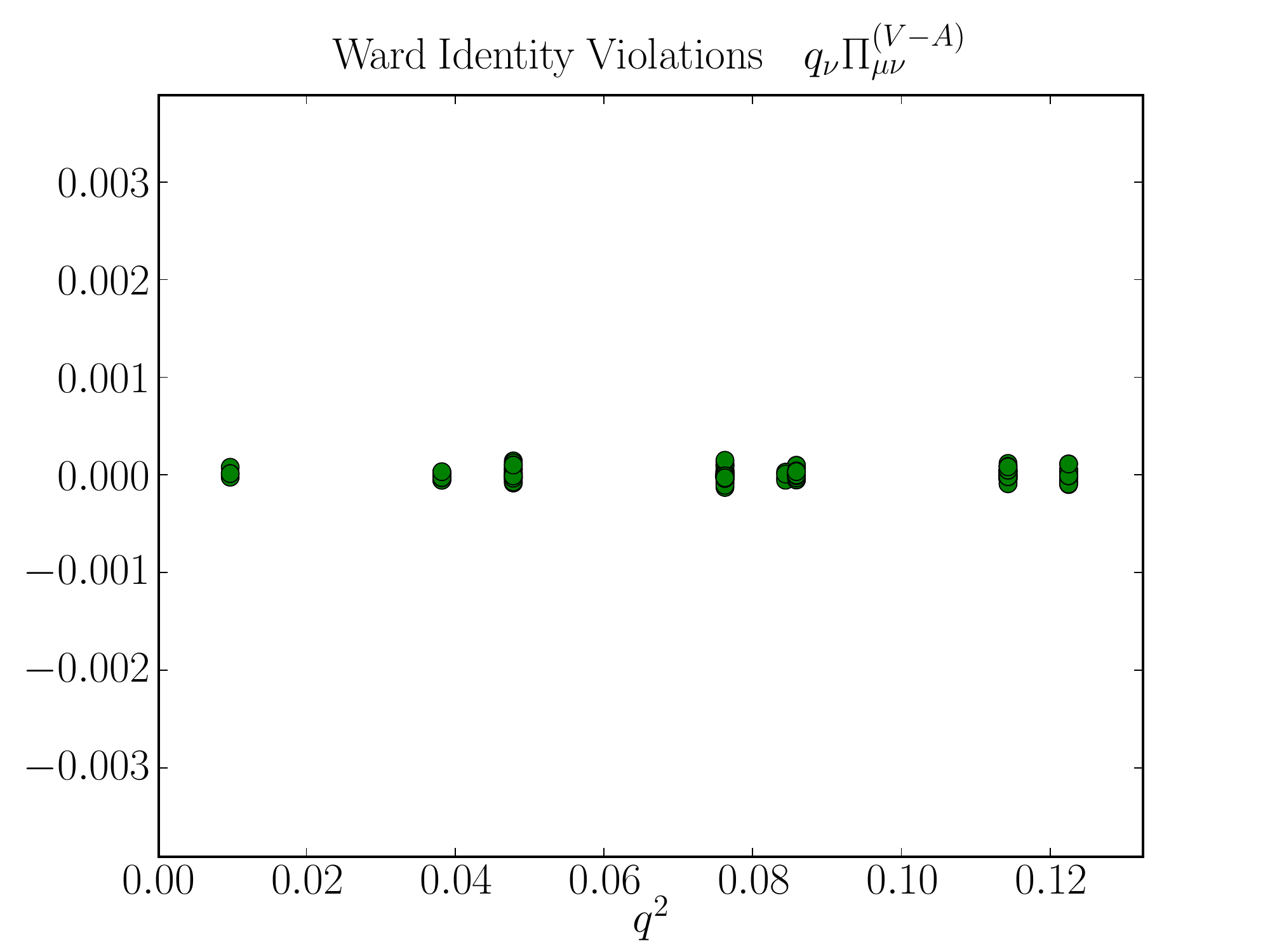}
  \caption{Ward identity violations in the chiral limit for the vector
  and axial vector currents (left) and their difference (right). The
  same scale is chosen in both plots to visualise the
  cancellation.\label{fig:ward_id}}
\end{figure}
Contact terms in the Ward identities can yield finite contributions to
the Fourier transform, so that $\Pi^{V/A}_{\mu\nu}$ is no longer
transverse. However the contact terms cancel exactly in the difference
between the vector--vector and the axial--axial correlator for DWF in
the $L_s\to\infty$ and massless limit, provided the conserved/local
correlators are used. Any power--divergent contribution also cancels
in this difference as a result of the chiral symmetry of DWF.

Following the notation introduced in the previous section we test the
vector and axial Ward identities by extrapolating
$q_\nu\Pi^{V/A}_{\mu\nu}$ to the chiral limit. We find that Ward
identity violations are very similar for both currents and contact
term contributions are greatly suppressed in the difference
$\Pi_{\mu\nu}^{V-A}$, as shown in Figure~\ref{fig:ward_id}.  The
cancellation in the chiral limit also hints at only small effects from
the non-conservation of the axial current due to the residual mass of
Domain Wall Fermions.
\begin{figure}[b]
  \includegraphics[width=.45\textwidth]{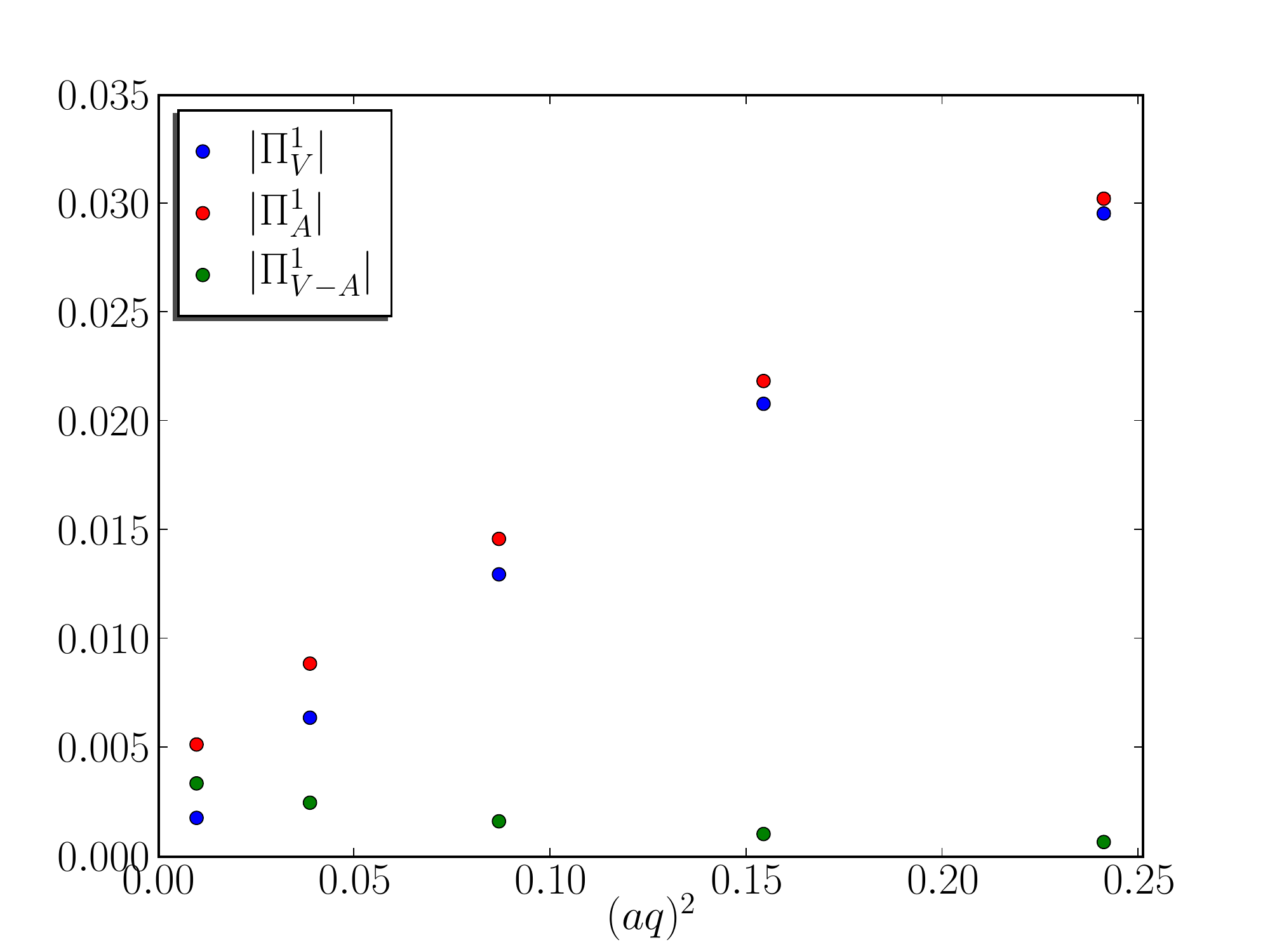}
  \includegraphics[width=.45\textwidth]{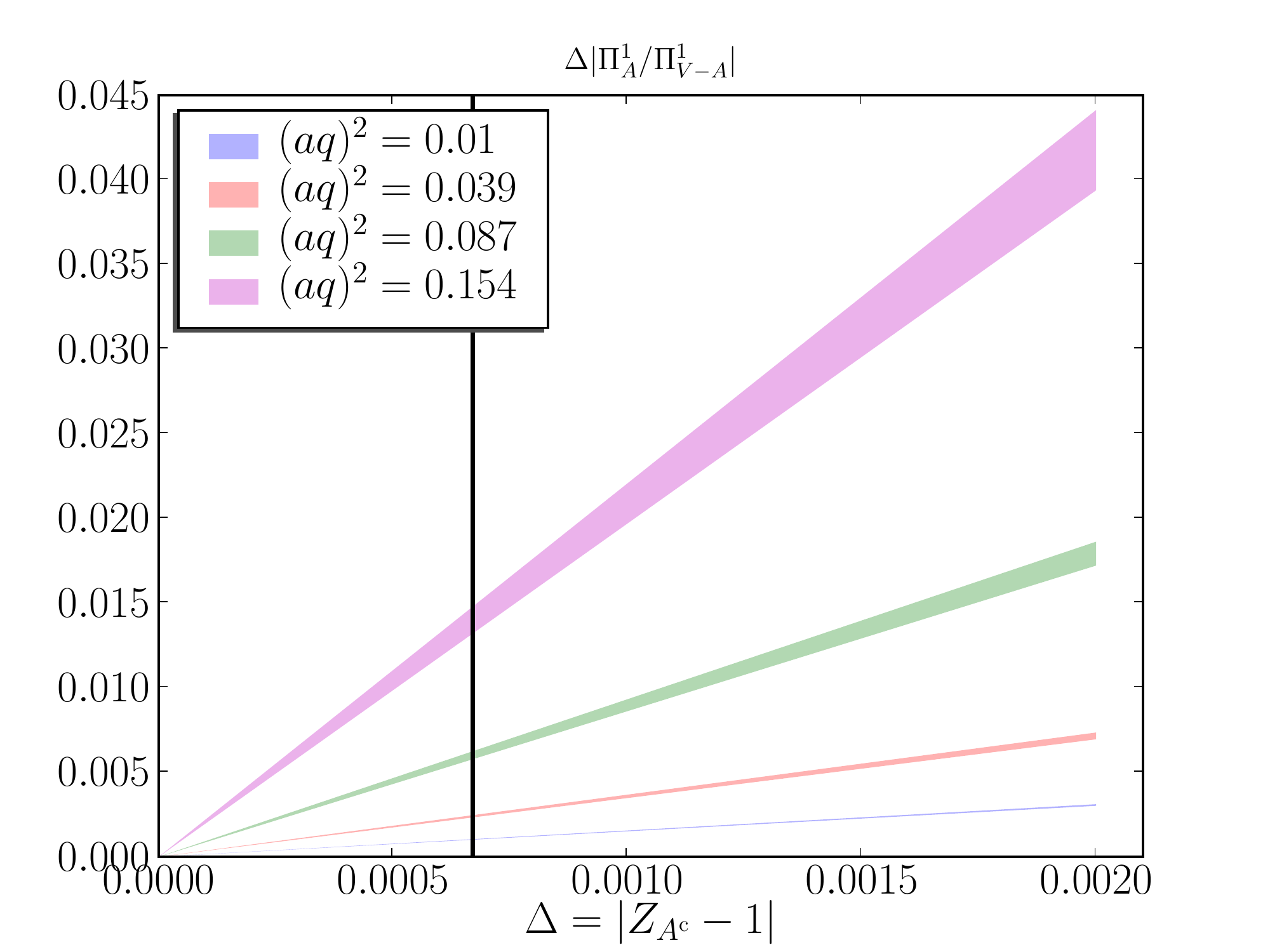}
  \caption{Left: cancellation between $\Pi^{(V)}$ and $\Pi^{(A)}$ as a
    function of $q^2$, Right: relative error for given $\Delta$ for
    the lowest momenta.\label{fig:zaerror}}
\end{figure}
The DWF axial current acquires a small multiplicative renormalisation
which vanishes in the $L_s\to\infty$ limit~\cite{Shamir:1993zy}. Since
our simulation is done at a fixed length in the fifth dimension
($L_s=16$) we have to consider this effect as part of our error
budget. In Refs.~\cite{Christ:2005xh,Sharpe:2007yd,Allton:2008pn} it
has been argued that $\Delta=|Z_{\mathcal{A}}-1|$ receives
contributions from both delocalized modes above the mobility edge
$\lambda_c$, and from localized near zero modes of $H_W$. The former
is proportional to $\e^{-\lambda_cL_s}$ while the latter is
proportional to $\frac{1}{L_s^2}$. These correspond to two different
contributions to the residual mass: the exponential piece is linear in
the corresponding component of $m_\text{res}$ while the latter is
quadratic in the localised contribution to $m_\text{res}$ that is
larger in our case.

As a pragmatic approach we vary $\Delta$ between 0 and
$3am_\text{res}$ and estimate the relative error on the difference
$\Pi^1$ as $\Delta \frac{\Pi^{(1),A}}{\Pi^{(1)}}$.  Our conclusion is
that there is no large cancellation for the local-conserved
correlators since $q^2\Pi^{(1),V}$ approaches zero, as shown in Figure
\ref{fig:zaerror}. We assume a conservative three percent systematic
error in $\Pi^{(1)}$ for the non-conservation of the axial current.

\section{Results}
\label{sec:res}
\begin{figure}[hbt]
\includegraphics[width=.7\textwidth]{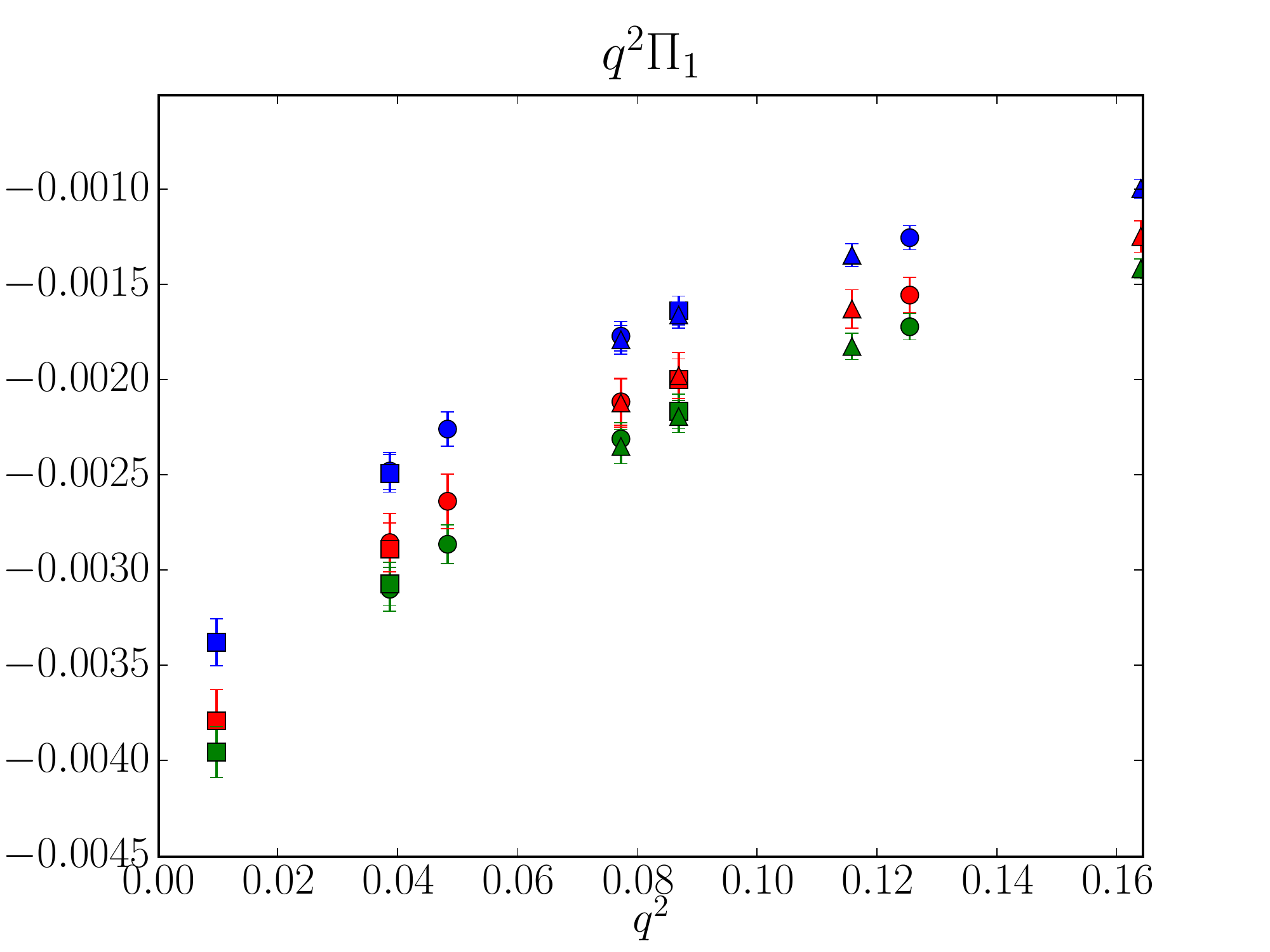}
\caption{$q^2\Pi^{(1)}$ at low momentum. The different symbols
  correspond to momenta of type $(n+1,0,0,0)$: squares, $(n,1,0,0)$:
  circles, $(n,1,1,0)$: triangles with $n=0,1,2$\label{fig:pi1}}
\end{figure}

The data presented is from the ensembles generated by the RBC and
UKQCD collaboration with the Iwasaki gauge action at $\beta=2.25$
which corresponds to a lattice spacing of
$a^{-1}=2.33(4)\,\text{GeV}$. The details of the ensembles will be
published in \cite{Aoki:2009xx}. We simulate with three values of the
light quark mass which correspond to pion masses on the range of
$290\,\text{MeV}\le m_\pi\le 400\,\text{MeV}$. Our chiral fits for $l_{5}$
rely on previous measurements of the pseudoscalar mass and decay
constant at the unitary quark masses. For convenience we summarize the
results obtained from correlators with gauge-fixed wall sources in
Table \ref{tab_meson} along with the vector and axial vector ground
state masses which are used as a consistency check using Weinberg's
sum rules and our data for $l_5$.
\begin{table}[hbt]
  \begin{center}
    \begin{tabular}{c|c|c|c|c}
      $am_l$ & $am_\pi$ & $af_\pi$ & $am_V$ & $am_A$ \\ \hline
      $0.004$ & $0.1269(4)$ & $0.0619(3)$ & $0.356(6)$ & $0.522(13)$ \\
      $0.006$ & $0.1512(3)$ & $0.0645(3)$ & $0.366(5)$ & $0.543(18)$ \\
      $0.008$ & $0.1727(4)$ & $0.0671(3)$ & $0.388(6)$ & $0.551(9)$ \\
    \end{tabular}
    \caption{Meson masses and pseudoscalar decay constant in lattice units.
      \label{tab_meson}}
  \end{center}
\end{table}
In our analysis of $\Pi^{(1)}$ we include spatial momenta up to
$(1,1,0)$ (or equivalent). We find excellent agreement between results
with and without spatial momentum for the local-conserved $\Pi^{(1)}$,
Figure \ref{fig:pi1}. This is in contrast to the local-local data
where there are two distinct branches of points. 
\begin{table}[ht]
  \begin{center}
    \begin{tabular}{c|c|c|c|c}
      momentum & $(ap)^2$ & $am$ & $q^2\Pi^{(1)}\cdot10^3$ & $L_{10}\cdot10^3$ \\ \hline

      $(1,0,0,0)$ & $ 0.010 $ & $ 0.004 $ & $ -3.37 ( 12 )$ & $ -5.6 ( 1.6 )$ \\
      & & $ 0.006 $ & $ -3.79 ( 16 )$ & $ -4.5 ( 2.1 )$ \\
      & & $ 0.008 $ & $ -3.95 ( 13 )$ & $ -7.0 ( 1.7 )$ \\ \hline
      $(2,0,0,0)$ & $ 0.039 $ & $ 0.004 $ & $ -2.49 ( 10 )$ & $ -4.02 ( 32 )$ \\
      & & $ 0.006 $ & $ -2.88 ( 14 )$ & $ -3.89 ( 45 )$ \\
      & & $ 0.008 $ & $ -3.07 ( 11 )$ & $ -4.55 ( 37 )$ \\ \hline
      $(0,1,0,0)$ & $ 0.039 $ & $ 0.004 $ & $ -2.5 ( 10 )$ & $ -3.99 ( 32 )$ \\
      & & $ 0.006 $ & $ -2.87 ( 15 )$ & $ -3.94 ( 50 )$ \\
      & & $ 0.008 $ & $ -3.22 ( 10 )$ & $ -4.06 ( 34 )$ \\ \hline
      $(1,1,0,0)$ & $ 0.048 $ & $ 0.004 $ & $ -2.25 ( 9 )$ & $ -3.77 ( 23 )$ \\
      & & $ 0.006 $ & $ -2.63 ( 14 )$ & $ -3.74 ( 37 )$ \\
      & & $ 0.008 $ & $ -2.86 ( 10 )$ & $ -4.17 ( 26 )$ \\ \hline
      $(0,1,1,0)$ & $ 0.077 $ & $ 0.004 $ & $ -1.77 ( 8 )$ & $ -3.07 ( 13 )$ \\
      & & $ 0.006 $ & $ -2.11 ( 12 )$ & $ -3.16 ( 20 )$ \\
      & & $ 0.008 $ & $ -2.31 ( 8 )$ & $ -3.51 ( 14 )$ \\
    \end{tabular}
    \caption{Results for $q^2\Pi^{(1)}$ and the resulting effective
    $L_{10}$ for the lowest momenta and all three masses. \label{tab_res}}
  \end{center}
\end{table}

As a first step we obtain effective values for $L_{10}$ for each mass
and momentum from our data for $\Pi^{(1)}$, Eq.(\ref{eq:pi1}). Our
results are summarised in Table \ref{tab_res}. Since ChPT is known to
have only a small radius of convergence in $p^2$, we restrict ourselves
to only the lowest momenta.  

\begin{figure}[hbt]
  \begin{center}
    \includegraphics[width=.7\textwidth]{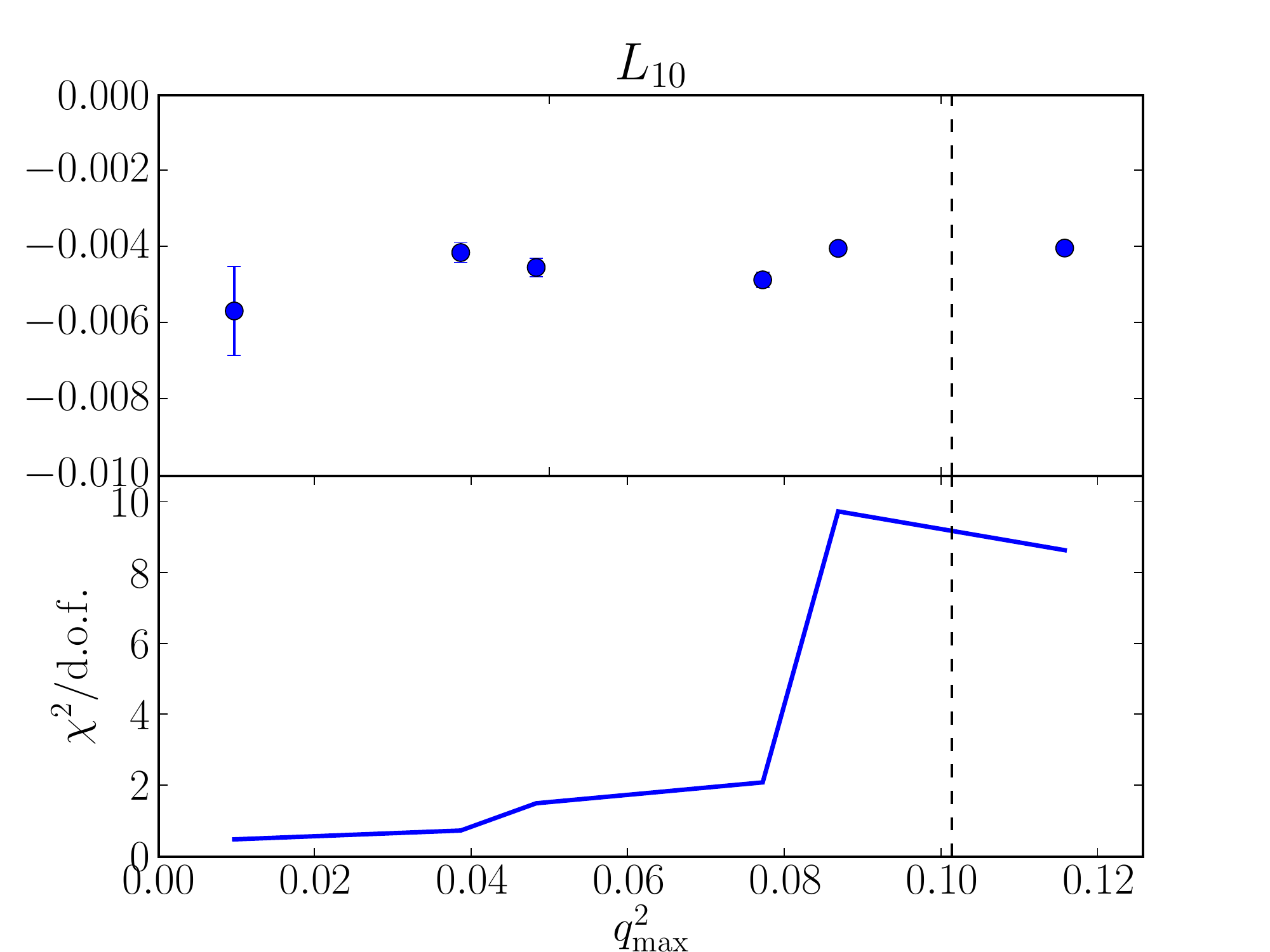}
    \caption{Upper panel: dependence of $L_{10}$ on the fit-range in
      $(ap)^2$. Lower panel: reduced $\chi^2$ of the corresponding
      fits described in the text. The dashed vertical line denotes the
      chiral scale in lattice units.\label{fig:l10_fitrange}}
  \end{center}
\end{figure}

\begin{figure}[hbt]
  \begin{center}
    \includegraphics[width=.45\textwidth]{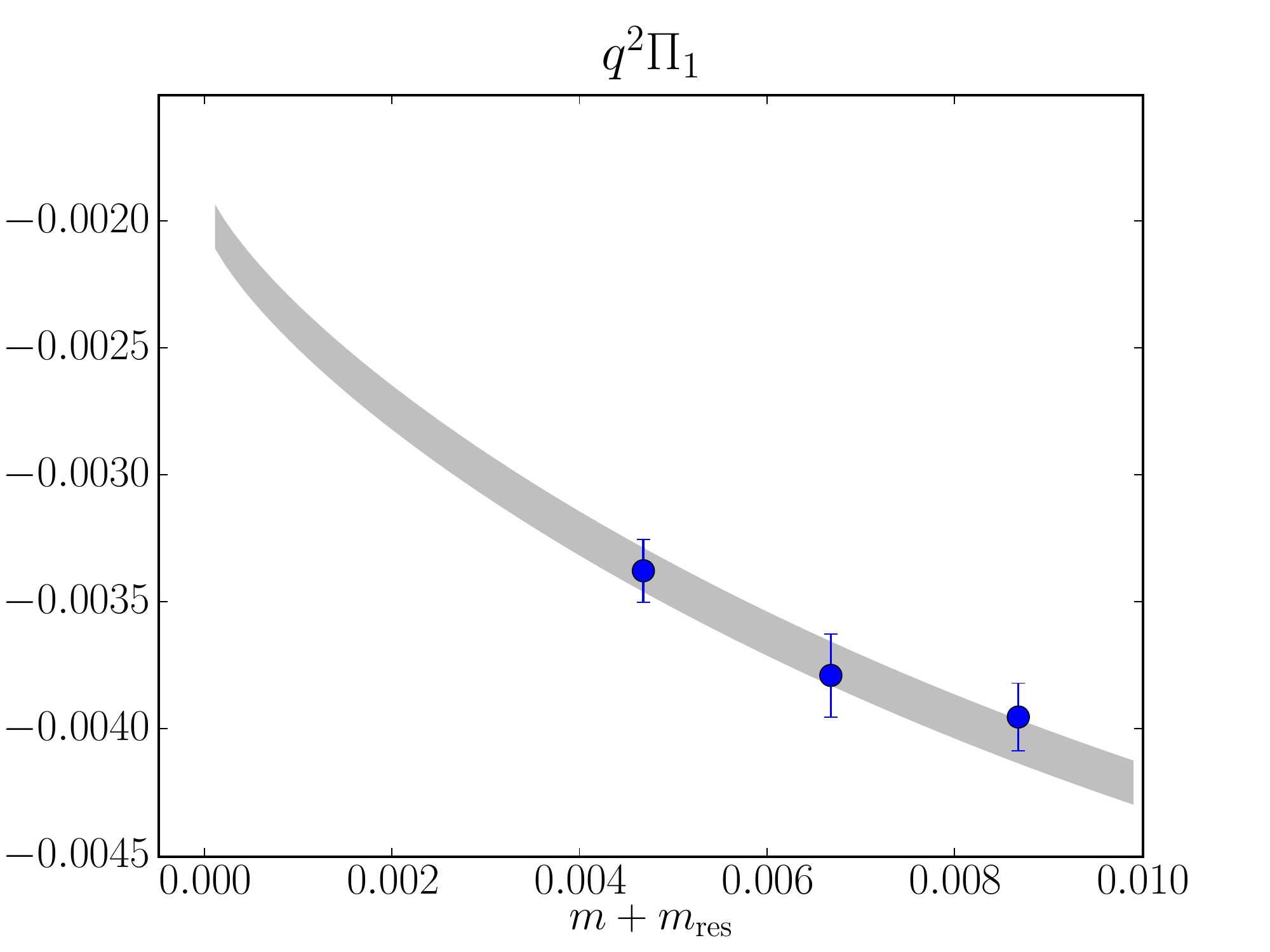}
    \includegraphics[width=.45\textwidth]{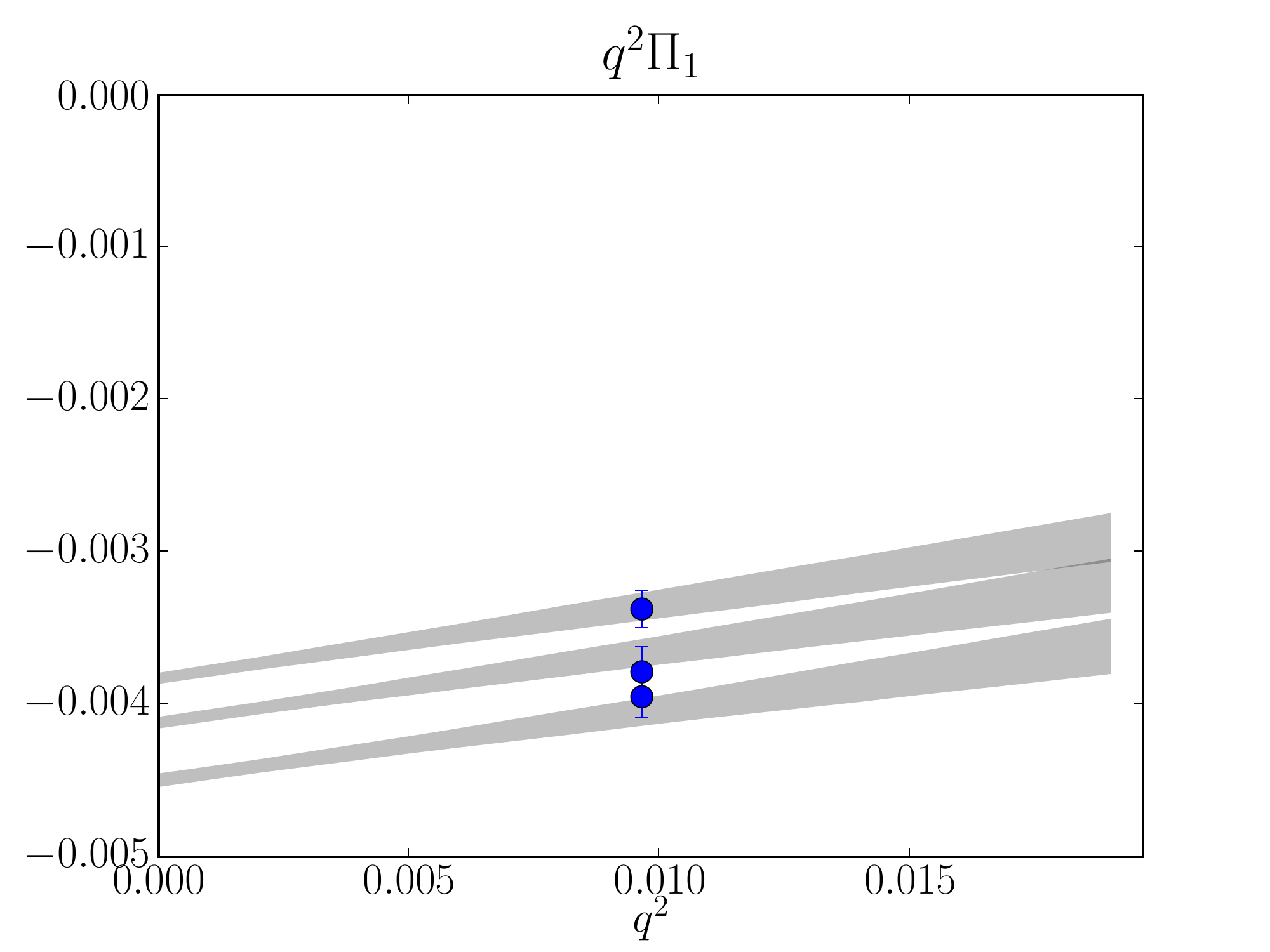}
    \caption{Data set and fit band used for the final result for $L_{10}$. 
      \label{fig:l10_bestfit}}
  \end{center}
\end{figure}

Now we perform a one parameter fit to obtain our central value for
$L_{10}$. In Figure \ref{fig:l10_fitrange} the dependence of $L_{10}$
on the momentum fit-range is shown for a fit including all three
masses. In the lower panel we observe that the $\chi^2/_\text{d.o.f.}$
becomes larger than one when more than the lowest two momenta are
included. Fits with fewer mass values exhibit the same behaviour. We
however conclude that including only the lowest momentum gives a more
reliable result since the higher momenta yield values for
$q^2\Pi^{(1)}$ which are consistently below the fit curve when
including only the smallest momentum, Figure
\ref{fig:l10_bestfit}. Our central value therefore is obtained from a
fit to all three masses and the lowest momentum only.
In Figure \ref{fig:l10_bestfit} the data points included in the determination of $L_{10}$
and the error band of the resulting fit is shown plotted against mass (left) and
momentum (right).
Our value for $L_{10}^r$ is
\begin{align}
  L_{10}^r(\mu=0.77\,\text{GeV})=-0.0057(11)_\text{stat}(7)_\text{sys}.
  \label{eq:result}
\end{align}
The systematic error has several contributions: fit-range, lattice
artefacts, scale setting, strange quark mass and finite volume, which
are described in more detail below.

We estimate the error on the chiral fit by varying the fit-range in
mass and $q^2$ and obtain an error of $0.0006$, or $11\%$. The leading
lattice artefacts with Domain Wall Fermions are parametrically of
order $a^2\Lambda_\text{QCD}^2$. We assume
$\Lambda_\text{QCD}=300\,\text{MeV}$, use $a^{-1}=2.33(4)\,\text{GeV}$
and double the result to obtain a three percent error. The uncertainty
in the lattice scale is relevant for the definition of the chiral
scale $\mu=m_\rho$. We vary $a^{-1}$ within its error and obtain an
additional two percent error. The dynamical strange quark mass in this
computation is fixed to $am=0.03$ which differs substantially from the
physical value. Using a reweighting procedure for the strange quark
determinant \cite{jung:2009xx} we vary the strange quark mass down to
the physical point in lattice units and find variations in $L_{10}^r$
of less than three percent. Possible finite volume effects (FVE) are
well under control, since at the lightest mass we have $m_\pi L=4$ and the
estimate for FVE for $f_\pi$ from \cite{Colangelo:2005gd} is
$0.5\%$. Adding the errors in quadrature we find a total systematic
error of $0.0007$ which is dominated by the error on the chiral fit.

\begin{figure}[hbt]
  \begin{center}
    \includegraphics[width=.7\textwidth]{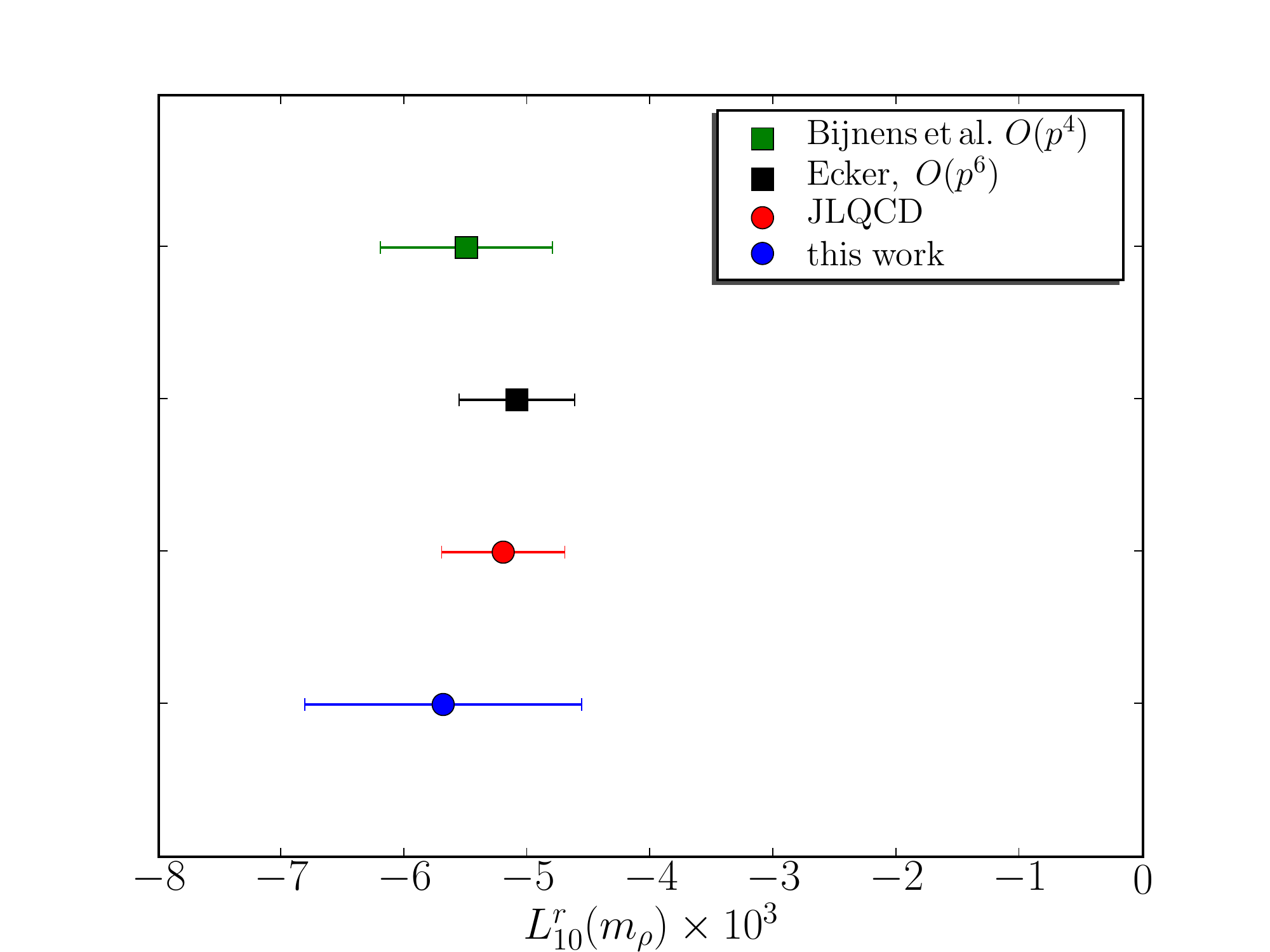}
    \caption{Comparison of $L_{10}^r$ with other determinations. 
      \label{fig:global}}
  \end{center}
\end{figure}
Comparing our result with the previous lattice determination
\cite{Shintani:2008qe} and phenomenological estimates
\cite{Bijnens:1994qh,Ecker:2007dj} we find it to be well consistent,
Figure \ref{fig:global}.

Lattice data for the lowest states in the vector and axial vector
channels allows us to crosscheck our result with the following sum
rules for the spectral densities $\rho_{V/A}$
\cite{Weinberg:1967kj,Das:1967ek},
\begin{subequations}
\begin{align}
  \int \dd s s\left(\rho_V(s)-\rho_A(s)\right) &= 0, \\
  \int \dd s \left(\rho_V(s)-\rho_A(s)\right) &= f_\pi^2, \\
  \int \dd s \frac{1}{s} \left(\rho_V(s)-\rho_A(s)\right) &= -8\overline{L}_{10}.
\end{align}
\end{subequations}
We assume that the spectral densities are saturated by the lightest resonances,
\begin{align}
\rho_{V/A}\approx f_{V/A}^2\delta(s-m_{V/A}^2).
\end{align}
The resulting simplified sum rules are solved for $\overline{L}_{10}$
using our values for $m_V,m_A$ and $f_\pi$ from Table \ref{tab_meson},
\begin{subequations}\label{eq:sr}
\begin{align}
  (f_Vm_V)^2-(f_Am_A)^2&=0, \label{eq:sr1}\\
  f_V^2-f_A^2 &= f_\pi^2, \label{eq:sr2} \\
  \frac{f_V^2}{m_V^2}-\frac{f_A^2}{m_A^2} &= -8\overline{L}_{10}.\label{eq:sr3}
\end{align}
\end{subequations}
\begin{figure}[hbt]
  \begin{center}
    \includegraphics[width=.7\textwidth]{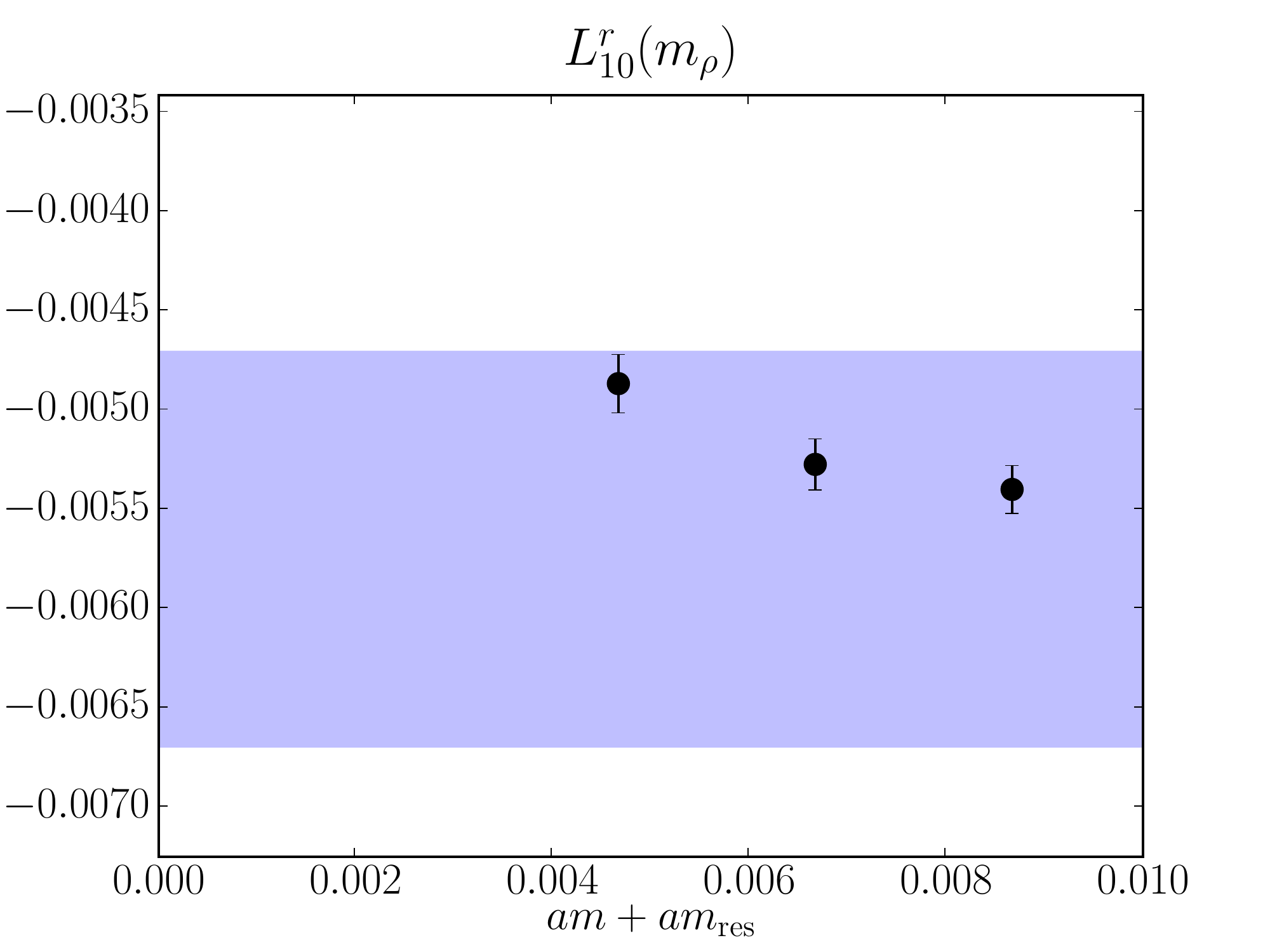}
    \caption{Results for $L_{10}^r$ from the sum rules in
    Eq. (\ref{eq:sr}) These are compared with the result from
    Eq. (\ref{eq:result}) which is indicated by the blue
    band.\label{fig:sumrules}}
  \end{center}
\end{figure}
Thus we obtain an independent estimate for $\overline{L}_{10}$ for
each quark mass which we convert to $L_{10}^r(m_\rho)$ using the pions
masses from Table \ref{tab_meson}. In Figure \ref{fig:sumrules}, these
results are shown by the black circles and are compared with our best
estimate (\ref{eq:result}), indicated in the figure by the blue band. We
find them to be well consistent. This agreement is a non-trivial
finding since it is found at finite lattice spacing and without a
chiral extrapolation for the values obtained from the sum rules
(\ref{eq:sr1}-\ref{eq:sr3}).

\section{Pion Mass Splitting}\label{sec:pion}
The computation of the vacuum polarisation functions over the complete
$q^2$ range allows the computation of the electromagnetic contribution
to mass splitting between the charged and neutral pions
\cite{Das:1967it},
\begin{align}
  m_{\pi^\pm}^2-m_{\pi^0}^2=-\frac{3\alpha}{4\pi}\int_0^\infty\dd
  q^2\frac{q^2\Pi^{(1)}}{f_\pi^2}=1261\,\text{MeV}^2.\label{eq:dgmly}
\end{align}
The simplest possible ansatz to describe the mass and momentum dependence
of $\Pi^{(1)}$ which respects the sum rules (\ref{eq:sr1},\ref{eq:sr2})
is
\begin{align}
  \begin{split}
    q^2\Pi^{(1)}(q^2,m)&=-f_\pi^2+\frac{f_V^2q^2}{m_V^2+q^2}-\frac{f_A^2q^2}{m_A^2+q^2},\\
    f_V&=x_1+x_2m_\pi^2,\quad m_V=x_3+x_4m_\pi^2,\\ 
    f_A^2&=f_V^2-f_\pi^2,\quad m_A=m_V\frac{f_V}{f_A},
  \end{split}\label{eq:ansatz}
\end{align}
where $x_1,\ldots,x_4$ are fit parameters.  We include all data
points up to $(aq)^2<1.0$ in the simultaneous fit to all three masses
and obtain a stable fit with
$\chi^2/_\text{d.o.f.}=1.04$. Extrapolating to the chiral limit we find
\begin{align}
  -\frac{3\alpha}{4\pi}\int_0^1\dd q^2 \frac{q^2\Pi^{(1)}}{f^2}=1040(220)
  \,\text{MeV}^2.\label{eq:intlow}
\end{align}
We have also tested the two different fit forms used in
\cite{Shintani:2008qe} which include an additional term, which is
relevant at high $q^2$, however we do not find a significant change in
the fit parameters $x_1,\ldots,x_4$ or a smaller
$\chi^2/_\text{d.o.f.}$ when we include these additional fit
parameters.
\begin{figure}[hbt]
  \begin{center}
    \includegraphics[width=.45\textwidth]{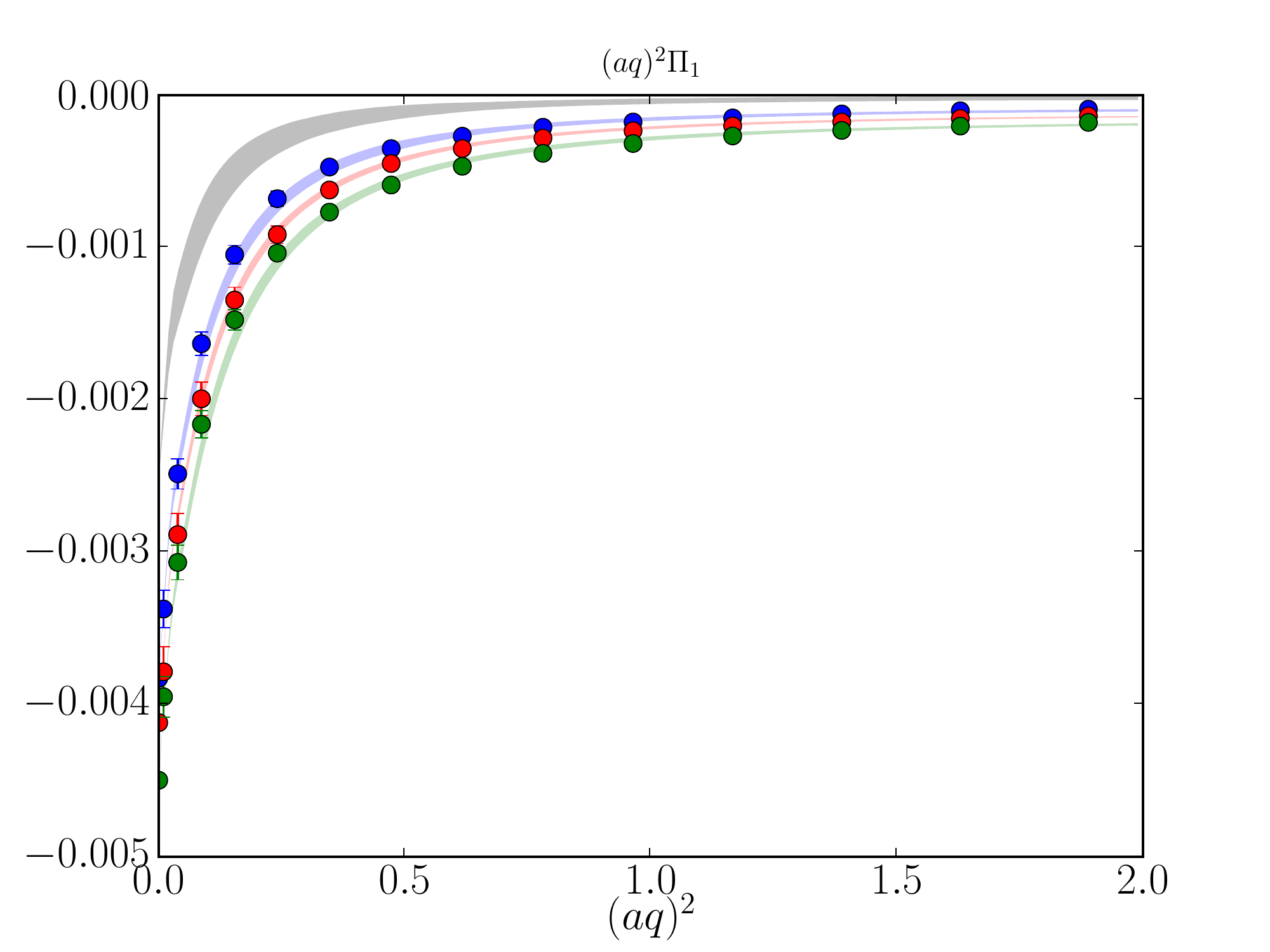}
    \includegraphics[width=.45\textwidth]{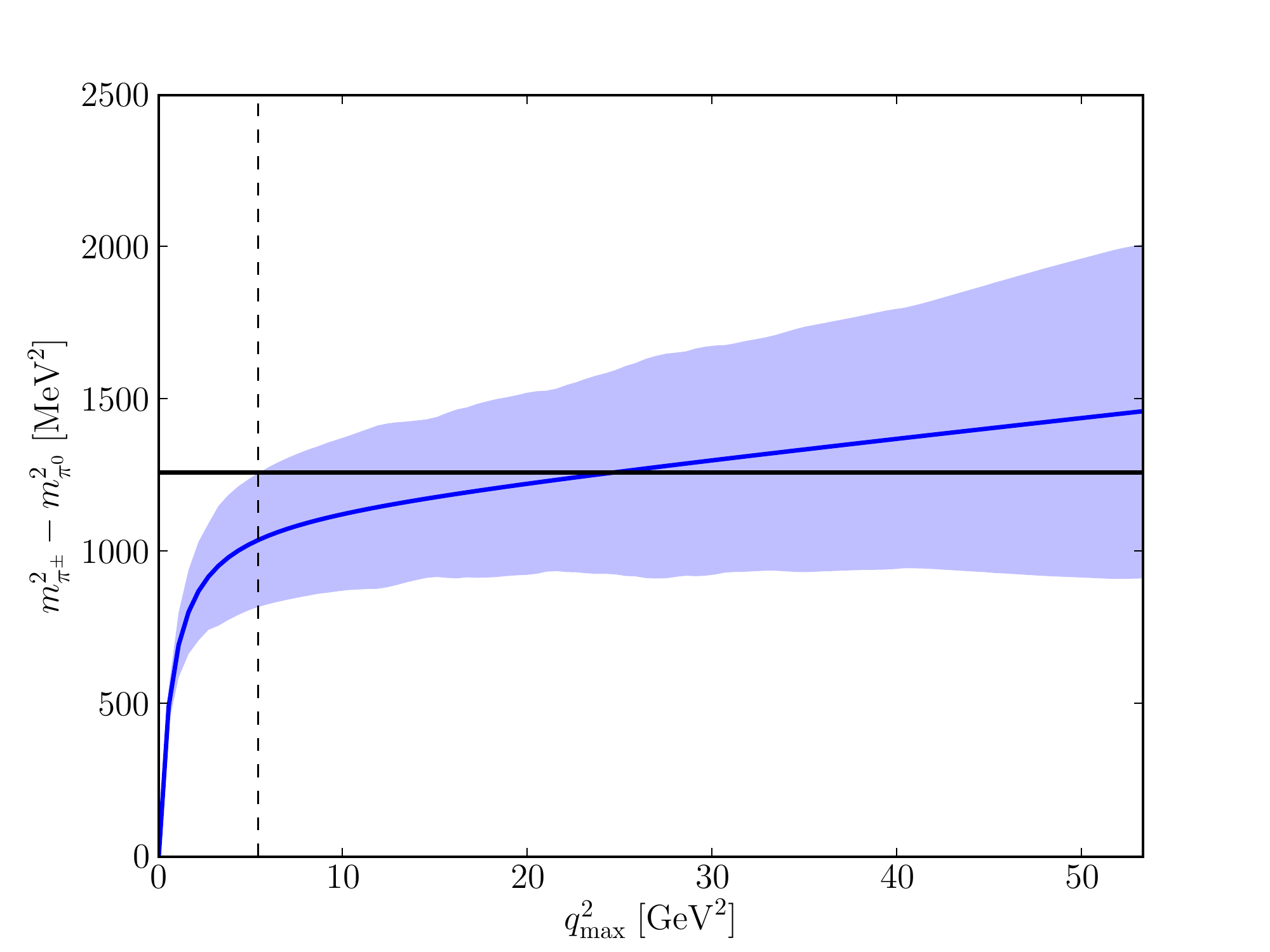}
    \caption{Left: fit with ansatz (\ref{eq:ansatz}), where the grey band
    denotes the result in the chiral limit. Right: mass splitting
    $m_{\pi^\pm}^2-m_{\pi^0}^2$ as a function of the cut-off momentum
    $q_\text{max}^2$, the horizontal line denotes the physical value,
    the dashed line marks our chosen cut-off.
      \label{fig:mpi}}
  \end{center}
\end{figure}
An illustration of the dependence on the cut-off is shown in Figure
\ref{fig:mpi}. The remaining part of the integral is estimated under
the assumption that the functional dependence follows the asymptotic
behaviour for large $q^2$,
\begin{align}
  \Pi^{(1)} \to \frac{c}{q^6}. \label{eq:highqsq}
\end{align}
Matching (\ref{eq:highqsq}) to our fit result in the chiral limit we
obtain $c=-0.0062(58)\,\text{GeV}^6$. Here we have doubled the error
to allow for possible deviations from the asymptotic form. The remaining
integral is then
\begin{align}
  -\frac{3\alpha}{4\pi}\int_1^\infty\dd q^2 \frac{q^2}{\Pi^{(1)}}{f^2}=140(130)
  \,\text{MeV}^2,
\end{align}
yielding a pion mass splitting of
\begin{align}
  m_{\pi^\pm}^2-m_{\pi^0}^2=1180(260)\,\text{MeV}^2.
\end{align}
This error does not include a rigorous investigation of the systematic
uncertainties as was done in Sec. \ref{sec:res} for $L_{10}$, however
we expect these uncertainties to be well within the large statistical
error of twenty percent obtained from the integral (\ref{eq:intlow}) with $0<(aq)^2\le1$. The
value we obtain is in excellent agreement with the experimental value
$m_{\pi^\pm}^2-m_{\pi^0}^2=1261\,\text{MeV}^2$~\cite{Amsler:2008zzb}.

Finally we can use the result of the fit defined in
Eq. (\ref{eq:ansatz}) to obtain another independent determination for
$L_{10}$. From the slope of the ansatz (\ref{eq:ansatz}) in the chiral
limit, we find $L_{10}^r(m_\rho)=5.19(14)\cdot10^{-3}$ which is in
agreement with our ChPT analysis. This is a nice check of the
saturation of the Weinberg sum rules. Unfortunately the large error
on our determination of $l_5$ does not allow a more quantitative
comparison.

\section{Conclusions}
\label{sec:conc}

We have computed the $S$ parameter in QCD using the gauge configurations
generated by the RBC-UKQCD collaboration for 2+1 flavors of dynamical
DWF fermions. According to the procedure outlined in
Ref.~\cite{Shintani:2008qe}, the $S$ parameter is extracted from the
form factor that appears in the parametrization of the VV-AA
correlator. Our final result is 
\begin{equation}
  \label{eq:finalres}
    L_{10}^r(\mu=0.77\,\text{GeV})=-0.0057(11)_\text{stat}(7)_\text{sys};
\end{equation}
where the error is still dominated by the statistical precision of our
simulation. On the other hand, the systematic error is dominated by
the choice of the fit-range used in the chiral extrapolation. A
further improvement to these simulations would be to include partially
twisted boundary conditions, enabling access to smaller momenta where
ChPT can be employed reliably. Using results from meson spectroscopy,
we were able to check the saturation of the Weinberg sum rule by the
lowest--lying resonances in QCD. At the current level of accuracy, we
find that the contribution from the lowest--lying resonances accounts
for the total value of the $S$ parameter, which confirms a
widely--used assumption in phenomenological studies. Our best estimate
for the $S$ parameter (\ref{eq:sl5}) with a Higgs boson mass of
$m_H=120\,\text{GeV}$ is
\begin{equation}
  \label{eq:spar}
    S=0.42(7),
\end{equation}
where we have rescaled the renormalization scale $\mu$ with the ratio
of the pion decay constant to the Higgs vacuum expectation value
$v=246\,\text{GeV}$.

From the same correlators we have extracted the electromagnetic pion
mass splitting, and the result $\Delta m_\pi^2=1180\pm
260\,\text{MeV}^2$ is in excellent agreement with the experimental
results. Despite the large error, the present computation shows that
it is possible to extract this quantity from the currently available
DWF ensembles.

The results presented here are only made possible through the chiral
properties of DWF which ensure the cancellation of the
power--divergent contributions in the current correlators. The QCD
analysis we have performed lays the groundwork for the computation of
the $S$ parameter in potential Technicolor model candidates such as
those proposed Refs.~\cite{Foadi:2007ue,Ryttov:2008xe}. Computing the
$S$ parameter in Technicolor theories is a crucial step in identifying the
models that survive the constraints from electroweak precision
measurements. We plan to extend this computation to the models
recently simulated in Refs.~\cite{DelDebbio:2008zf}.

\section*{Acknowlegments}
We would like to thank Tom Blum, Norman Christ and Roger Horsley for stimulating
discussions. LDD and JZ are supported by Advanced STFC fellowships
under the grants PP/C504927/1 and ST/F009658/1. PAB is supported by an
RCUK fellowship.

\bibliography{spar}
\end{document}